# A Theory of Visibility Measures in the Dissociation Paradigm

**Thomas Schmidt & Melanie Biafora**

University of Kaiserslautern, Germany

Corresponding author:

Thomas Schmidt
University of Kaiserslautern, Faculty of Social Sciences, Experimental Psychology Unit
Erwin-Schrödinger-Str. Geb. 57
D-67663 Kaiserslautern, Germany
E-Mail: thomas.schmidt@sowi.uni-kl.de

## Abstract

**Research on perception without awareness primarily relies on the dissociation paradigm, which compares a measure of awareness of a critical stimulus (direct measure) with a measure indicating that the stimulus has been processed at all (indirect measure). We argue that dissociations between direct and indirect measures can only be demonstrated with respect to the *critical stimulus feature* that generates the indirect effect, and the observer's awareness of that feature, the *critical cue*. We expand Kahneman's (1968) concept of *criterion content* to comprise the set of all cues that an observer actually uses to perform the direct task. Different direct measures can then be compared by studying the overlap of their criterion contents and their containment of the critical cue. Because objective and subjective measures may integrate different sets of cues, one measure generally cannot replace the other without sacrificing important information. Using a simple mathematical formalization, we redefine and clarify the concepts of validity, exclusiveness, and exhaustiveness in the dissociation paradigm, show how dissociations among different awareness measures falsify simple theories of "consciousness", and formulate the demand that theories of visual awareness should be sufficiently specific to explain dissociations among different facets of awareness.**

**Keywords:** perception without awareness, dissociation paradigm, double dissociations, perceptual awareness scale, theories of consciousness

## Introduction

Research on perception without awareness relies primarily on the *dissociation paradigm*, which compares two types of measurement (Reingold & Merikle, 1988). *Indirect measures* are used as indicators that a critical stimulus has been processed in the first place (e.g., a masked prime or a binocularly suppressed image). Typical indicators are priming effects in response times. *Direct measures* are supposed to measure visual awareness for the critical stimulus that provoked the indirect effect. Typical such measures are discrimination accuracy or visibility ratings. [1]

Historically, most researchers have aimed for a *simple dissociation* between direct and indirect measures, which is observed when the indirect measure shows a clear nonzero effect while the direct measure indicates null sensitivity (T. Schmidt & Vorberg, 2006). Time and again, the dissociation paradigm has been attacked for seldom if ever demonstrating a simple dissociation convincingly (Eriksen, 1960; Holender, 1986; Meyen, Zerweck, Amado, von Luxburg, & Franz, 2020), even though rather convincing demonstrations of simple dissociations exist (e.g., Norman, Akins, Heywood, & Kentridge, 2014; F. Schmidt & T. Schmidt, 2010; Vorberg et al., 2003). In response to this problem, a minority of papers have aimed for a *double dissociation* pattern (Albrecht, Klapötke, & Mattler, 2010; Biafora & T. Schmidt, 2020; Lau & Passingham, 2007; Maniscalco, Peters, & Lau, 2016; Mattler, 2003; Merikle & Joordens, 1997; Vorberg et al., 2003). A double dissociation occurs when an experimental manipulation leads to an *increase* in performance in the indirect measure but a *decrease* in performance in the direct measure, or vice versa: for instance, an increase in priming effects over experimental conditions accompanied by a decrease in discrimination accuracy for the prime (Vorberg et al., 2003). Double dissociations are more powerful than simple ones because they do not require null sensitivity in the direct measure while also operating under milder measurement assumptions. They indicate that direct and indirect measures cannot both be monotonic functions of a single source of (conscious) information (T. Schmidt & Vorberg, 2006; T. Schmidt, 2007).

The way awareness of the critical stimulus should be measured is a matter of debate. Two types of measures can be distinguished (Seth, Dienes, Cleeremans, Overgaard, & Pessoa, 2008). *Objective measures* are responses to the critical stimulus that can be compared with the actual stimulus characteristics (e.g., its color or shape) and are therefore classifiable as correct or incorrect (e.g., yes-no detection or discrimination; two-alternative forced choice; recognition; identification). *Subjective measures* are reports of an internal state that cannot be validated externally (e.g., ratings of stimulus brightness, clarity of impression, or confidence in correct identification; Cheesman & Merikle, 1984, 1986; Reingold, 2004). The distinction between subjective and objective measures thus refers to the task mode rather than the content of the measure. Several authors argue that subjective and objective measures can be equally sensitive because they found that when participants report that subjective visibility is absent, their performance on an objective discrimination task was also at chance (e.g.,

Avneon & Lamy, 2018; Lamy, Alon, Carmel, & Shalev, 2015; Lamy, Carmel, & Peremen, 2017; Peremen & Lamy, 2014; Ramsøy & Overgaard, 2004). Other authors, however, have found marked differences in the data patterns from objective and subjective measures (e.g., Biafora & Schmidt, submitted; de Graaf, Goebel, & Sack, 2012; Jannati & DiLollo, 2012; Koster, Mattler, & Albrecht, 2020; Lau & Passingham, 2007).

In psychophysical procedures, objective and subjective measures are often used jointly, e.g., when constructing a receiver-operating characteristic (*ROC*) that plots objective hit and false alarm rates as a function of subjective confidence ratings. Signal Detection Theory (*SDT*; Green & Swets, 1966; Macmillan & Creelman, 2005) can be viewed as a model of the subjective experience of a stimulus when it is present or absent in unavoidable noise; it gives rise to objective performance if the observer applies a criterion to the subjective evidence that leads to discriminatory behavior. SDT thus gives room to subjective influences when separating sensitivity from response bias, and so do threshold-based models like the double high-threshold model (Malejka & Bröder, 2019). Nevertheless, some authors advocate the exclusive use of subjective measures, while others advocate the opposite, despite the close connection of the two in psychophysical theory.

### Itinerary for this paper

The purpose of this paper is to clarify the roles of indirect, objective, and subjective measures in the dissociation paradigm and draw conclusions for theories of visual awareness. We start by introducing the idea of a *critical cue*, the perceptual counterpart to the physical stimulus feature that generates the indirect effect, and argue that the critical cue constitutes an indispensable basis for any dissociation between direct and indirect measures. After showing that different direct measures can undergo surprising dissociations amongst each other, we extend an important idea in psychophysical research: the concept of *criterion content* (Kahneman, 1968) as consisting of a set of perceptual cues. This is the starting point for our Cue Set Theory (CST) of visibility measures in the dissociation paradigm. We explain how cues must be integrated to form measures of awareness, and use a simple mathematical formalization to redefine the concepts of exclusiveness, exhaustive reliability, and exhaustive validity of awareness measures. Next, we take some time to study different patterns of overlap in the criterion contents of objective and subjective measures, as well as their possible containment of the critical cue, and show that neither class of measures can generally replace the other without sacrificing crucial information. After briefly discussing the validity of the popular Perceptual Awareness Scale (PAS), we evaluate claims about measurement properties that are frequently evoked in the literature. In the final part of the paper, we return to the empirical fact that different measures of awareness of the same stimulus may undergo double dissociations among each other, meaning that one measure increases over experimental conditions while another decreases. We use T. Schmidt and Vorberg's (2006) mathematical methods to prove three propositions: 1) that double dissociations among direct measures imply that they cannot all depend monotonically on the same single source of information ("no single source for double dissociations"); 2) that a simple theory that explains awareness in terms of a single monotonic process cannot explain a double dissociation between two direct measures ("no simple theory for double dissociations"); 3) that any theory of awareness that seeks to avoid being falsified in this way needs to explain the entire set of awareness measures, including the dissociations ("explaining the gradient"). We end the paper with a proposal to advance a more modest view of visual awareness and unconscious perception: Instead of trying for sweeping theories of "consciousness", we advocate studying task dissociations not only between direct and indirect measures, but also among different direct measures, and to build theories that are sufficiently specific to address the many differences between all those facets of conscious and unconscious vision. Without loss of generality, we focus on the domain of visual perception, but note that our theory can be extended to other sense modalities as well as to fields like implicit memory and learning, implicit decision making, and others.

A few words about the role and purpose of mathematical formalization in this paper. Our use of elementary mathematics (limited to basic set theory and the simple algebraic concept of monotonicity of functions) is not intended to flabbergast readers with complicated expressions for ideas that are already commonplace in consciousness science. Rather, we are trying to *pinpoint* those ideas by transforming them into clearly defined concepts that are specific enough to carry a mathematical proof. Doing this has three important advantages. First, a more explicit formulation reveals the scopes and limits of those concepts and how they are related to each other. Second, it gives critical readers the chance to examine the exact assumptions underlying our arguments, to attack our basic tenets by questioning the assumptions, and to arrive at new tenets by using alternative assumptions. Third, it helps prevent using important methodological concepts in a fuzzy, metaphorical way.

**Criterion content and the critical feature**

The concept of criterion content was introduced by Daniel Kahneman in his studies on *metacontrast*, a form of visual backward masking (Breitmeyer & Öğmen, 2006). In his famous review paper (Kahneman, 1968), he argues that participants in psychophysical experiments may use sources of information quite different from what the researchers expect (also see Hake, Faust, McIntyre, & Murray, 1967, for an early quantitative approach to this problem). For instance, when asked to discriminate whether a masked prime is a square or a diamond, a participant may develop a strategy to monitor a particular spot on the screen, inferring that the prime was a square whenever she detects a flicker in that spot. That participant may successfully perform the task without ever consciously seeing the prime's shape: Her criterion content is based on flicker at a specific location, not on perceived shape. Kahneman stresses that to examine an observer's criterion content, it is necessary to consider the phenomenology of the observations: "[…] a fuller description of the code that the subject uses in mapping his private experience onto responses to the experimenter's questions" (Kahneman, 1968, p. 410). Let's examine this concept a little further.

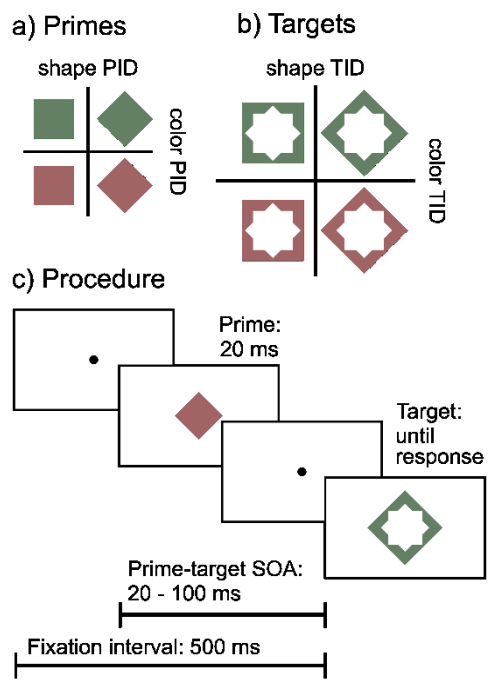

Fig. 1: *A hypothetical experiment. a, b) Primes and targets differ in two dimensions, color or shape, and the target serves to mask the prime by metacontrast. c) In two target identification tasks, participants respond either to the shape or to the color of the target, and priming effects from different prime types serve as indirect measures of shape or color processing, respectively. But how should we measure the visibility of the primes?*

Figure 1 shows the outline of a hypothetical response priming experiment where a prime is followed by a target at various stimulus-onset asynchronies. Because the inner contours of the target are adjacent to the prime contours, the target also serves as a metacontrast mask of the prime and can strongly reduce its visibility (provided that the colors are sufficiently desaturated; T. Schmidt, 2000). Primes and targets are red squares, green squares, red diamonds, and green diamonds. In two *target identification tasks* (*TIDs*, performed in different sessions), participants give a speeded response either to the shape of the target (*shape TID*) or to its color (*color TID*). In shape TID, the shape of the prime will activate the correct or incorrect response, resulting in longer response times when prime shape and target shape are inconsistent than when they are consistent. This priming effect is our indirect measure indicating processing of prime shape. In color TID, it is the color that will prime responses to the target, and that priming effect is an indirect measure indicating processing of prime color. Previous studies show that in two-dimensional stimuli and separate TID tasks like this, it is only the task-relevant feature that primes the response while the task-irrelevant feature does not affect response times, even though the stimulus material is identical in both tasks (e.g., Heinecke, 2000; Seydell-Greenwald & T. Schmidt, 2012; Tapia, Breitmeyer, & Shooner, 2010). What is important here is that even though both tasks use identical stimuli, the *critical feature* is different for each task. We define the critical feature as the physical stimulus distinction that drives the indirect effect -- the difference between square and diamond primes in shape TID, and the difference between red and green primes in color TID. In other words, the critical feature is always implied by the processing requirements of the indirect task. This is crucial for the logic of the dissociation paradigm: Any dissociation between an indirect and a direct measure (be it objective or subjective) is only meaningful when the direct task measures awareness of the critical feature -- otherwise there is a mismatch between the tasks (*D-I mismatch*; T. Schmidt & Vorberg, 2006). In our example experiment, shape TID thus requires a direct task asking for shape, and color TID requires a direct task asking for color. In contrast, a detection instead of a discrimination task would fail to match either indirect task because the priming effect is driven by the shape or color of the prime, not its presence or absence (Reingold & Merikle, 1988).[2]

Both objective and subjective tasks can be used to measure awareness of the critical feature. An objective measure could directly ask the observer to indicate whether the prime was, for instance, red or green. A subjective measure could ask, "Rate the clarity with which you perceived the color of the prime". Both questions clearly address the critical feature, but only the objective one explicitly asks about its identity and can be compared with the actual stimulus.

## Dissociations among multiple direct measures

The classical dissociation paradigm is usually discussed in terms of one indirect and one direct variable. However, when several direct measures are employed in the same experiment, surprising dissociations can occur among them. Lau and Passingham (2007) used masked squares and diamonds under metacontrast masking and compared an objective direct measure (percentage of correct discriminations) with a subjective one (percentage of "seen" ratings). They showed that subjective ratings could still differ when objective performance was equated. In the same vein, Sackur (2013) showed participants pairs of metacontrast events at different target-mask SOAs and asked them to rate their subjective similarity. He then used multidimensional scaling to argue that even if two metacontrast conditions lead to the same objective discrimination performance, their subjective appearance can still differ. Vorberg et al. (2003) presented participants with arrow primes masked by metacontrast and showed that while the ability to detect the prime increased with prime-mask SOA, the ability to discriminate the prime's pointing direction (which was the critical feature that generated the priming effect in a companion task) remained at chance.

Most recently, Koster, Mattler, and Albrecht (2020) further explored the possibilities of employing multiple direct measures. They presented square or diamond-shaped primes (shown for 24 ms) that were followed by square- or diamond-shaped masks (shown for 108 ms). The prime-mask SOA was varied parametrically, ranging from 24 to 84 ms (much like our example experiment shown in Fig. 1, but with black shape stimuli on white background). Metacontrast masking gives rise to a rich phenomenology of subjective percepts that depend on stimulus factors (timing, contrast, eccentricity, shape, relative energy), but also vary strongly between observers (Albrecht et al., 2010; Albrecht & Mattler, 2010, 2012, 2016). In the first part of their study, the authors collected detailed verbal descriptions of what the observers experienced in the different experimental conditions. From these reports, they derived seven subjective direct measures to be used in the second experiment. In that experiment, participants were presented with all experimental conditions (2 primes x 2 targets x 6 SOAs) for six sessions (following an entire additional session as practice). On each trial, participants indicated by a yes/no decision whether one particular percept had occurred (a subjective task). There was also an objective direct task in which participants tried to discriminate whether the masked prime was a square or diamond. There was no indirect task. The results offer a singularly rich picture of the subjective experience of 24 well-trained observers, measured with high precision.

First of all, the objective prime discrimination measure (in $d'$ units) showed that for most observers, performance was either a declining or u-shaped function of SOA (the phenomenon of "type-B masking" that meta-contrast is famous for; Breitmeyer & Öğmen, 2006; Kahneman, 1968). Only two observers showed an increase in performance with SOA, and two more observers performed at chance level throughout. Averaged across all observers, however, performance was declining with SOA and leveled off at $d' \approx 0.5$ (which is low performance but clearly above chance).

Each of the subjective measures showed a similarly distinctive pattern, but often very different from objective performance. The likelihoods of (1) *perceiving a prime before the mask*, of (2) *perceiving the prime as dark*, and of (3) *perceiving no prime at all* were distinctly u-shaped and markedly increased at longer SOAs, while the likelihood of (4) *perceiving a bright prime* was constant with SOA. Interestingly, the likelihood of (5) *perceiving rotation between the prime and mask* increased with SOA, but only when prime and mask were inconsistent in shape (none of the other measures showed this dependence on prime-target consistency). Only the likelihood of (6) *perceiving the prime as filling out the mask* and of (7) *perceiving an expansion from prime to mask* had the same declining time-course as the objective measure.

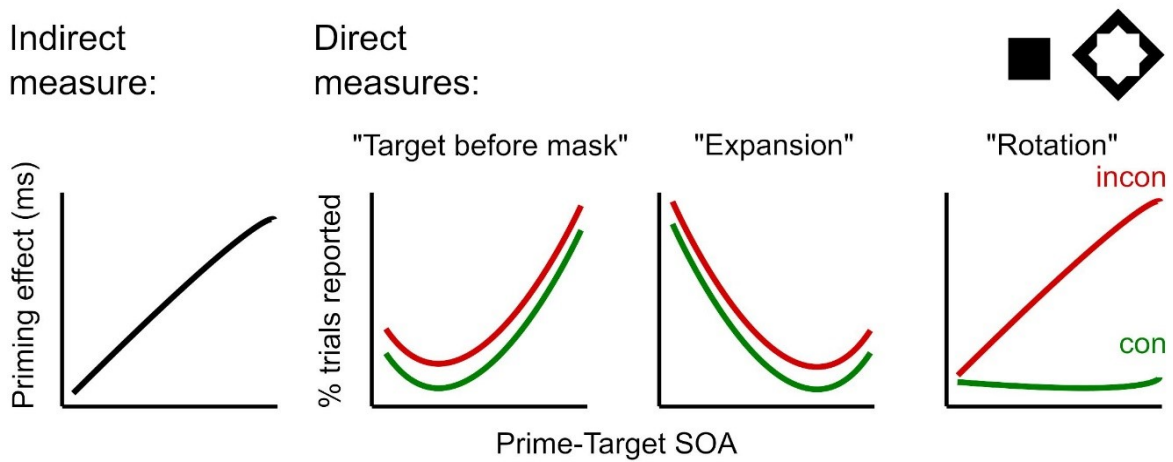

Fig. 2: *Hypothetical data inspired by Koster et al.'s (2020) study (there were additional awareness measures in their study). While a priming effect in response times (indirect measure) monotonically increases with prime-target SOA, three subjective direct measures of awareness for the prime show a variety of patterns. The measure "target before mask" increases, while the measure "expansion" decreases. The behavior of the third measure, "rotation", depends on prime-target consistency: it increases in inconsistent trials only.*

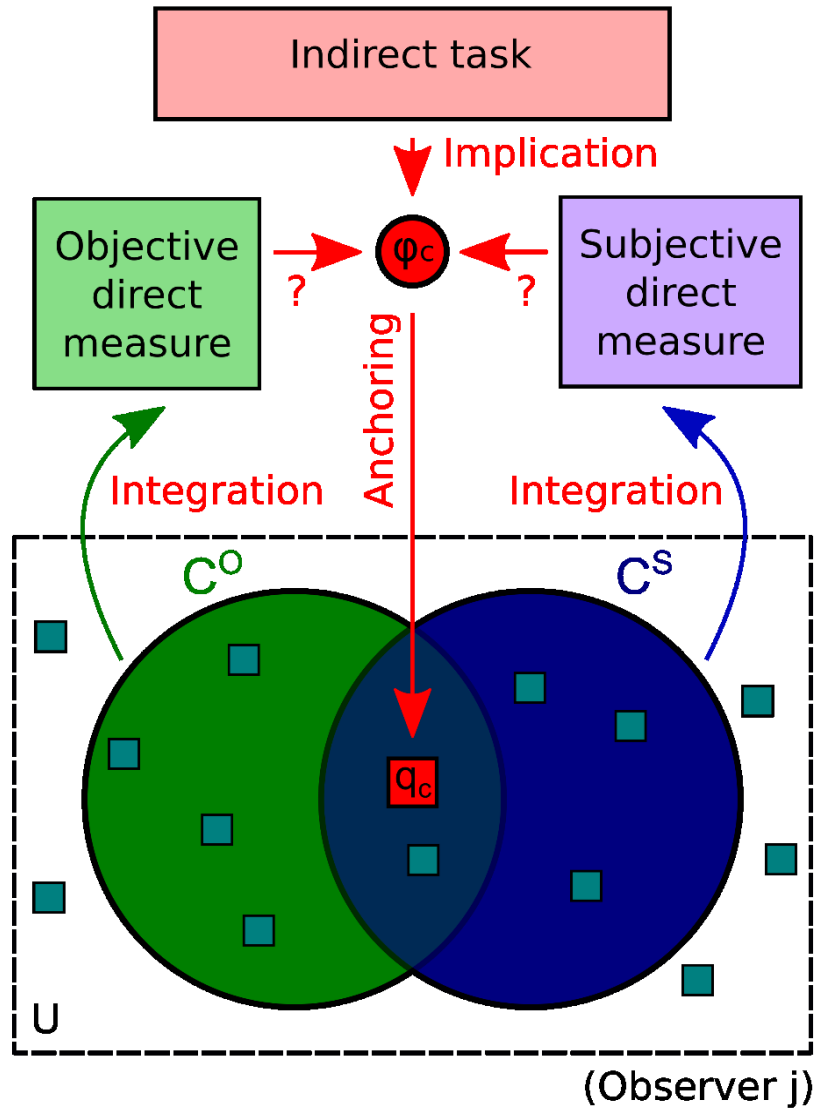

*Fig. 3: Outline of Cue Set Theory (CST). Relations between one objective and one subjective measure in the dissociation paradigm. The indirect task implies the critical feature, $\varphi_c$, which is the stimulus feature that generates the effect in the indirect measure (e.g., priming of responses by shape or by color, respectively). Dissociations between direct and indirect tasks can only be demonstrated on the basis of the critical feature, otherwise there is a mismatch between tasks. The critical feature thus provides an anchor for the critical cue, $q_c$, which is defined as perceptual awareness of the critical feature. Objective and subjective direct measures are performed on the basis of their respective criterion contents, $C^O$ and $C^S$, which are observer- and task-specific. Criterion contents are the sets of cues (shown as small squares) factually used to perform the respective task. Cues can be of diverse origin and need not be perceptual; they must be integrated to form the direct measures. Some of them may remain unused by either task (forming set U). Methodological debates revolve around the question whether or not direct tasks measure awareness of the critical feature (indicated by question marks). -- Note that this depiction only shows a special case where the criterion contents are partially overlapping and the critical cue is contained in both of them.*

**Criterion content as a set of cues**

With this example of dissociated direct measures in mind, we are now ready to expand on Kahneman's (1968) concept of criterion content. We do that by redefining criterion content as a set of cues that an observer uses to perform the task.

While the critical feature ($\varphi_c$) is defined on the basis of stimulus differences (e.g., in physical features, category membership, or whichever distinction is driving the indirect effect), criterion content is based on sources of information (*cues*) within the cognitive system of an observer that are used to perform the direct task. Cues can be based on diverse sources of information. The *critical cue ($q_c$)* is the one that directly corresponds to visual awareness of the critical feature (e.g., awareness of prime shape in shape TID, awareness of prime color in color TID). But the critical cue is not necessarily what is factually used by the participant. What cues beside the critical cue can be used by a participant trying to perform the direct task?

First of all, *(1) auxiliary cues* are perceptual cues other than the critical cue that can be helpful in discriminating the prime, like a perceived flicker, a brightening or darkening, an expansion or rotation (Albrecht et al., 2010). In addition, *(2) sensorimotor cues* might arise from the response conflict induced by the prime, which is known to activate an initial motor response that can easily lead to a response error when the prime is inconsistent (Panis & T. Schmidt, 2016; T. Schmidt, 2000; T. Schmidt, Niehaus, & Nagel, 2006; T. Schmidt & F. Schmidt, 2009; F. Schmidt, Weber, & T. Schmidt, 2010; Vorberg et al., 2003). If target ID and prime ID are performed on the same trial, observers may be able to monitor the speed and accuracy of their response, the perceived effort, or the initial tendency to respond. Such cues are especially useful when direct and indirect measures are employed on the same trial: for instance, an error in the indirect task may lead the observer to infer that the prime was inconsistent with the target, enabling an informed guess of its identity (Biafora & T. Schmidt, 2020). Similarly, *(3) decisional cues* result from the perceptual decision process (measured, for instance, by confidence ratings or by type-II $d'$ – a measure of how well observers are able to classify their perceptual decisions as correct or incorrect; Maniscalco & Lau, 2012; Zehetleitner & Rausch, 2013). *(4) Fringe cues* refer to hunches, gut feelings, fringe sensations, or other exotic sources of information. *(5) Strategic cues* are not experiential in nature but still might aid or harm performance in the direct task. They are not related to the stimulus or response but stem from prior knowledge (or assumptions) about the task. For instance, participants may use prior information about the relative frequency of different primes or congruency conditions (educated guessing), or they might try to count different prime types and then pick the one that is more frequent (or less frequent, if the participant assumes that primes are drawn without replacement). -- Finally, some authors worry that the direct measure could be contaminated by *(7) automatic cues*: The prime could activate its associated response not only in the indirect but also in the direct task (Kiesel, Wagener, Kunde, Hoffmann, Fallgatter, & Stöcker, 2006), which would lead to an overestimation of its visibility.

Now we are ready to expand Kahneman's notion of criterion content:

> *Definitions (i).* Let $q$ be a cue, and let $q^T_{ij}$ be cue $i$ that participant $j$ factually uses to perform task $T$. Then, the discrete and finite set $C^T_j$ forms the *criterion content* for this participant and task, $C^T_j = \{q \mid q = q^T_{ij}\}$. Cues that are not part of any criterion content are in the set $U_j$ of unused cues, $U_j = \{q \mid q \notin C^T_j$ for all $T\}$. Objective direct measures, $D^O_j$, and subjective direct measures, $D^S_j$, are functions defined on their respective criterion contents, $D^O_j \equiv f^O_j(C^O_j)$, $D^S_j \equiv f^S_j(C^S_j)$.

In what follows, we will usually drop the $j$ subscript to simplify notation, keeping in mind that $C^T$, $D^T$, $f^T$, and $U$ are always observer-specific.

Note that our definitions allow objective and subjective measures to be based on *identical* criterion contents. It is a frequent misunderstanding that "subjective" content is best captured by a "subjective" measure, as if objective measures were somehow void of subjective content. The difference between objective and subjective measures is solely in whether the observer's responses can be compared with the external stimulus, not on the nature of the internal evidence on which they are based.

## Direct measures integrate the cues in their criterion content

We just defined direct measures as functions of the criterion content; we now outline possible functions. We assume that in order to perform a psychophysical task, observers have to *integrate* the cues in their task-specific criterion content (Anderson, 1992; Marks & Algom, 1998). Generally, integration can be accomplished in many ways given that cues can differ in their scaling properties (e.g., they may form indicator, ordinal, interval, or ratio scales; they may also be vector-valued, like color coordinates). For the sake of illustration, let us assume that all the cues in the criterion content are real-valued random variables coded such that larger values denote more evidence for the information addressed by the cue, and that an observer is trying to maximize her performance in an objective direct task (for typical experiments, that means maximizing response accuracy in identifying the critical feature). How this is done in an optimal way that maximizes the reliability of the integrated measure is a classical problem in mathematical statistics (Cochran, 1937). Under the assumption that the cues are uncorrelated, it is optimal to weigh the cues according to their reliabilities, i.e., their ability to predict the critical feature (e.g., Drewing & Ernst, 2006; Landy, Maloney, Johnston, & Young, 1995; Oruç, Maloney, & Landy, 2003).

But optimizing a measure's reliability is only one way of integrating the cues in a criterion content; there could be radically different criteria for integration. In the psychological literature on heuristic decision making, many integration schemes are discussed (Gigerenzer & Gassmeier, 2011). For example, an observer may have many cues available to her, but choose to restrict her entire criterion content to only one cue (tantamount to setting its weight to 1 and all others to 0). If this is the critical cue, we call this measure *exclusive for the critical cue.* This property may not guarantee that the criterion content allows for optimal performance (because additional cues may have led to further improvement), but it means that the critical cue is the sole basis of performance and that the measure is free of contamination by other sources of information.

While objective measures may be optimized with respect to objective performance, no such external criterion exists for subjective measures. Subjective measures could aim to optimize internal criteria instead, like confidence in a decision (Locke, Landy, & Mamassian, 2022; Zehetleitner & Rausch, 2013). An example would be an observer who has no perceptual cues available but a subjective feeling whether or not his decision was correct. Occasionally, observers in masked prime discrimination tasks report monitoring their initial motor impulses (they "follow where their finger wants to go"; see Kiesel et al., 2006). These observers seem to optimize the perceived difference between the initial motor impulse and the subsequent discrimination response -- we privately call this the "Zen Mode" of Prime ID. Still other observers might be content with a measure that keeps the task comfortable and minimizes the perceived effort invested (e.g., the occasional negligent subject who always presses the same key). As a result, subjective measures can vary a lot in the range of awareness levels they can cover, as well as how they respond to a state of unawareness (Wierzchoń, Asanowicz, Paulewicz, & Cleeremans, 2012).

Ultimately, performance in a direct task is determined both by the specific cues in the task's criterion content, $C^t$, and by the manner of their integration, $f^t(C^t)$. It is therefore difficult to say whether there is any optimal set of cues or any optimal integration function for a given observer, because it is possible that a given combination of criterion content and integration function might be outperformed by some other combination. We also have to deal with the possibility that an observer may integrate a criterion content in such a twisted way that the measure changes sign with respect to the information provided by a cue, such that the measure decreases when the evidence in question actually increases. The following definitions will help to exclude such cases from further consideration.

> *Definitions (ii).* Assume a measure $M$ with criterion content $C^M = \{q_1, q_2...\}$ and all cues $q_i$ coded such that larger values indicate stronger evidence. $M$ is a *monotonic integrator* of $C^M$ if for any cue $q_i$ and all other cues remaining equal, $q_i' \geq q_i$ implies $M(..., q_i', ...) \geq M(..., q_i, ...)$. $M$ is an *exhaustive integrator* of $C^M$ if strict inequalities hold, such that for any cue $q_i$ and all other cues remaining equal, $q_i' > q_i$ implies $M(..., q_i', ...) > M(..., q_i, ...)$.

Whether or not an integrator is exhaustive is an all-or-none property; it makes no sense to state that one measure is more exhaustive than another. We will see shortly that the strict inequalities required for exhaustive integrators lead to the classical "exhaustiveness problem" of the dissociation paradigm (Reingold & Merikle, 1988). Note that the properties of monotonic and exhaustive integration only need to hold in the long run at the level of expected values. Also note that a measure that does not respond to changes in its criterion content at all is already sufficient to satisfy the requirement of monotonic integration (e.g., a participant who always gives the same response). Monotonic integration is only violated when a measure systematically (i.e., in the long run) responds in reverse to the cues in its criterion content. We shall assume throughout the paper that all direct measures considered are monotonic integrators.

The concepts of monotonic and exhaustive integrators also link the present paper to the previous one by T. Schmidt and Vorberg (2006). In that paper, we discuss three types of dissociations between direct and indirect measures: single, double, and sensitivity dissociations (the latter occur when the indirect measure outperforms the direct one; cf. Meyen et al., 2020). Apart from describing the data patterns, we also investigate the measurement assumptions necessary for interpreting those dissociations as evidence for unconscious cognition. Those assumptions all concern the question whether a direct or indirect measure is an exhaustive or merely a monotonic function of conscious information, $c$, or unconscious information, $u$. In the present framework, we can allow $c$ and $u$ to be based on multiple sources of conscious or unconscious information. Any assumptions of exhaustiveness or monotonicity of direct and indirect measures in T. Schmidt and Vorberg (2006) can then be replaced by assumptions of exhaustive integration or monotonic integration, respectively.

**A new look at exhaustiveness and exclusiveness**

Reingold and Merikle (1988) argue that a direct measure of visual awareness in the dissociation paradigm should have two properties. First, it should be *exhaustive* for visual awareness, which means that all relevant aspects of visual awareness are covered by the measure. This is a logical requirement for interpreting simple dissociations: Zero sensitivity or chance performance in the direct measure can only imply the absence of awareness if it is certain that no aspects of awareness escape measurement. Second, the direct measure should be *exclusive* for visual awareness. Even though this is not a logical requirement for the dissociation paradigm (see the mathematical appendix in T. Schmidt & Vorberg, 2006, where the exclusiveness assumption is never needed), it is a desirable property because a non-exclusive direct measure could be contaminated by unconscious information (e.g., automatic cues; Kiesel et al., 2006). Cue Set Theory, the reformulation of criterion content as a set of cues that must be integrated to form perceptual measures, allows us to formulate these properties more specifically.

Actually, exhaustiveness turns out to have two aspects: the *validity* of a direct measure's criterion content (i.e., the choice of cues in it), and the *reliability* of the direct measure after the cues have been integrated (cf. Shanks & St. John's distinction between information criteria and sensitivity criteria for direct measures). The validity requirement for exhaustiveness is that all relevant cues need to be part of the criterion content -- we call such a measure *exhaustively valid*. First and foremost, this usually concerns the critical cue, but may also include some auxiliary perceptual, sensorimotor, decisional, or fringe cues. Whenever the criterion content fails to include a cue that could help predict the critical feature, it is possible that this cue is just the one that generates nonzero performance in the indirect measure. It would then be possible that the direct measure shows null sensitivity only because it misses this crucial cue. -- Note that what constitutes a "relevant" cue is a matter of the substantial research question: In contexts like masked priming, it may be desirable that the direct measure contains only perceptual cues, while in contexts like intuitive decision making decisional and fringe cues are of theoretical interest.

The reliability aspect of exhaustiveness was introduced by Reingold & Merikle (1988) and further investigated by T. Schmidt and Vorberg (2006) in their study of different types of dissociation between direct and indirect measures. They showed that the assumption of exhaustiveness postulates a psychophysical measure that is a strictly monotonic function of conscious information (here, of all relevant cues in the criterion content). Strict monotonicity means that the measure is able to detect any increase in conscious information, however small, like an infinitely sensitive barometer that never "hangs". In other words, the measure must be an exhaustive integrator, and only then can we call it *exhaustively reliable*. In contrast, a measure that is merely a monotonic integrator but not an exhaustive one may fail to detect an actual increase in awareness: there may be some cue that increases in value without the direct measure picking it up.

Importantly, a measure can only be exhaustively reliable if it is exhaustively valid; therefore, like exhaustive validity, exhaustive reliability depends on the research context. The problem with exhaustive reliability is that such a measure would have to be noise-free and infinitely sensitive -- clearly an untenable assumption in a psychophysical context. Fortunately, T. Schmidt and Vorberg (2006) show that only simple dissociations require an exhaustiveness assumption while double dissociations do not. Therefore, if the indirect measure increases under experimental manipulation while the direct measure decreases, the direct measure is no longer required to show zero sensitivity, and integrators only need to be monotonic, not exhaustive.

Like exhaustiveness, exclusiveness can also be redefined under Cue Set Theory. We call a direct measure *exclusive for relevant cues* when its criterion content only includes theoretically relevant cues and excludes all others. This requirement is met when $C^T \subseteq R^{T*}$, where $R^{T*}$ is the set of all cues relevant in task $T$. A special case applies when a measure is *exclusive for the critical cue*, i.e., when $C^T = \{q_c\}$ for some task $T$. Such a measure not only needs to include the critical cue, but also to exclude all others. We can summarize the foregoing in the following definitions, keeping in mind that we have to distinguish between theoretically relevant cues and additional, irrelevant ones:

*Definitions (iii).* Let $R^{T*} = \{r_1, r_2...\}$ be the set of all cues that are theoretically relevant for a specific task $T$. A criterion content that only consists of the critical cue, $C^T = \{q_c\}$, is *exclusive for the critical cue*. A criterion content that only consists of relevant cues, $C^T \subseteq R^{T*}$, is *exclusive for relevant cues*. A measure $C^T$ is *exhaustively valid* if its criterion content contains all relevant cues, $R^{T*} \subseteq C^T$.

Assume a direct measure $D^T$ with a criterion content $C^T$ exclusive for relevant cues and with all its cues coded such that larger values indicate stronger evidence. Let $R^T \subseteq R^{T*}$ denote the set of relevant cues in the criterion content. Let $S^T$ be the set of all remaining cues in the criterion content, so that $C^T$ consists of a set of relevant and of a set of irrelevant cues, $C^T = \{R^T, S^T\} = \{..., r_i, ...; ..., s_i...\}$. Then the measure is said to be *exhaustively reliable* if it is (i) exhaustively valid, and (ii) an exhaustive integrator of the relevant cues in its criterion content, so that, all other cues remaining equal, $r_i' > r_i$ implies $D^T(..., r_i', ...; ..., s_i, ...) > D^T(..., r_i, ...; ..., s_i, ...)$ for all $r_i, s_i \in C^T$.

Observe the interesting logical relationship between exclusiveness and exhaustiveness: If a criterion content is exclusive for relevant cues, $C^T \subseteq R^{T*}$; if it is exhaustively valid, $R^{T*} \subseteq C^T$; and if it has both properties, $C^T = R^{T*}$. Note that our definition allows for additional, irrelevant cues $s_i$ as long as they do not spoil the monotonic integration of the relevant cues $r_i$.

Again, the strict inequality in the definition means that an exhaustively reliable measure will respond to any change, however small, in the relevant cues in its criterion content, which guarantees that a nonresponse of the measure implies a nonresponse in the relevant cues. Again, which cues are "theoretically relevant" is a question of the substantial research paradigm, but if the goal is to establish a dissociation from an indirect measure, one of the $r_i$ must be the critical cue, $q_c$.

## Choosing direct measures that capture the critical cue

We can now apply this classification to concrete measures. For instance, are there measures that are exclusive for the critical cue, i.e., respond only to $q_c$ but to nothing else? To be a plausible candidate for this remarkable property, such a measure will have to ask directly for the critical feature, because it is the one that drives the indirect effect. It is usually easy to formulate such candidate measures, both in their objective and their subjective variants. For example, in the color task of our example experiment it is the difference in prime color (red or green) that defines the priming effect. This implies that the objective task most likely to be exclusive for the critical cue would be discrimination of the prime as red or green, of course using the same stimuli as the indirect task. A corresponding subjective measure would ask whether the observer *perceived* the prime as red or green, but in a way that does not force the participant to pick any one color (because this would turn the measure into an objective one). Some possibilities are the following: "Could you see whether the prime was red or green? Answer yes or no"; "Please rate how clearly you saw that the prime had one color rather than the other". Note that those questions ask specifically about the distinction between the two colors, but that the answers cannot be classified as correct or incorrect. Admittedly, these examples of subjective measures seem contrived, which is why we see the value of subjective measures mostly when it comes to facets of awareness not directly based on the critical feature.

A very interesting sort of direct measure is a bipolar rating scale, with one pole marked as "I clearly saw that the prime was red" and the other pole as "I clearly saw that the prime was green", with the various degrees of clarity in between. This is a *hybrid measure* that unites an objective and a subjective measurement: Choosing the "red" or "green" half of the scale is an objective discrimination task using criterion content $C^O$, and choosing the magnitude of the rating is a subjective task using criterion content $C^S$. If the wording is specific enough, this is an attractive candidate for a task that is both an exclusive objective task and an exclusive subjective task, in which case $C^O = C^S = \{q_c\}$. The popular bipolar scale that uses confidence instead of clarity ratings is of course another example of a hybrid measure, but one where $C^O$ and $C^S$ are hoped to give independent rather than concordant pieces of information.

The problem is that in order to know that a task is exclusive for the critical cue, we have to exclude the possibility that any cue except the critical one influences the behavior of the observer. Given the large number and idiosyncratic nature of possible task strategies, this is practically impossible. However, observers can be trained to adjust their criterion content (or its manner of integration) as desired by the experimenters (Koster, Mattler, & Albrecht, 2020).

**Overlap between criterion contents**

Our set-theoretic representation now allows us to compare objective and subjective measures with criterion contents $C^O$, $C^S$, that are both intended to measure awareness in a dissociation paradigm. We have seen that the indirect task defines a critical feature, and therefore the paradigm requires at least one direct measure that includes the critical cue to avoid mismatch between direct and indirect tasks. The deciding questions becomes: where is the critical cue in relation to the criterion contents? And if an objective and a subjective task are employed in tandem, what are their respective roles in a possible dissociation? We have already treated the case that the criterion contents $C^O$, $C^S$ of an objective and a subjective measure are identical. Now we discuss the remaining scenarios (Fig. 4).

*Fig. 4a, b:* If the criterion content for the subjective measure completely includes the criterion content for the objective measure, $C^O \subseteq C^S$, $C^S$ is said to *cover* $C^O$. In that case, all the cues that could be used to optimize the objective measure are also available for the subjective measure (Fig. 4a). As a special case, both measures may use the same cues, $C^O = C^S$. Another special case occurs if the criterion content of the objective measure is a proper subset of the subjective one, $C^O \subset C^S$, in which case we say that $C^S$ *outmatches* $C^O$. When this occurs, the subjective measure can use all the cues available for the objective measure, but not vice versa. In other words, there are cues being used that are *unique* to the subjective measure (Fig. 4b). Subjective measures are interesting precisely because there are uniquely subjective cues, such as stimulus clarity or decision confidence, that have no counterpart in objective measures.

*Fig. 4c, d:* These concepts apply symmetrically for objective and subjective measures. In Fig. 4c, $C^O$ covers $C^S$, and all the cues that could optimize the subjective measure are also available for the objective measure. In Fig. 4d, $C^O$ outmatches $C^S$: the objective measure can use all the cues that could optimize the subjective measure, but not vice versa. Examples of objective measures that rely on cues unique to them are same-different tasks where the difference between stimulus conditions is difficult to verbalize, the study of differential behavior in animals who cannot provide subjective measures, or the observation of differential sucking rates in infants.

*Fig. 4e:* Each criterion content has elements that are not included in the other one, $C^O \setminus C^S \neq \varnothing$ AND $C^S \setminus C^O \neq \varnothing$. In this case, $C^O$ and $C^S$ do not cover each other. Each of them uses cues that are unique for the measure, and neither can outmatch the other.

*Fig. 4f:* If the two criterion contents have no cue in common, $C^S \cap C^O = \varnothing$, they are *disjoint* and the tasks are performed on the basis of entirely different sources of information. In that case, the subjective measure cannot use any of the cues that could optimize performance in the objective measure, and vice versa. Of course, disjoint criterion contents imply that $C^S$ does not cover $C^O$ and vice versa.

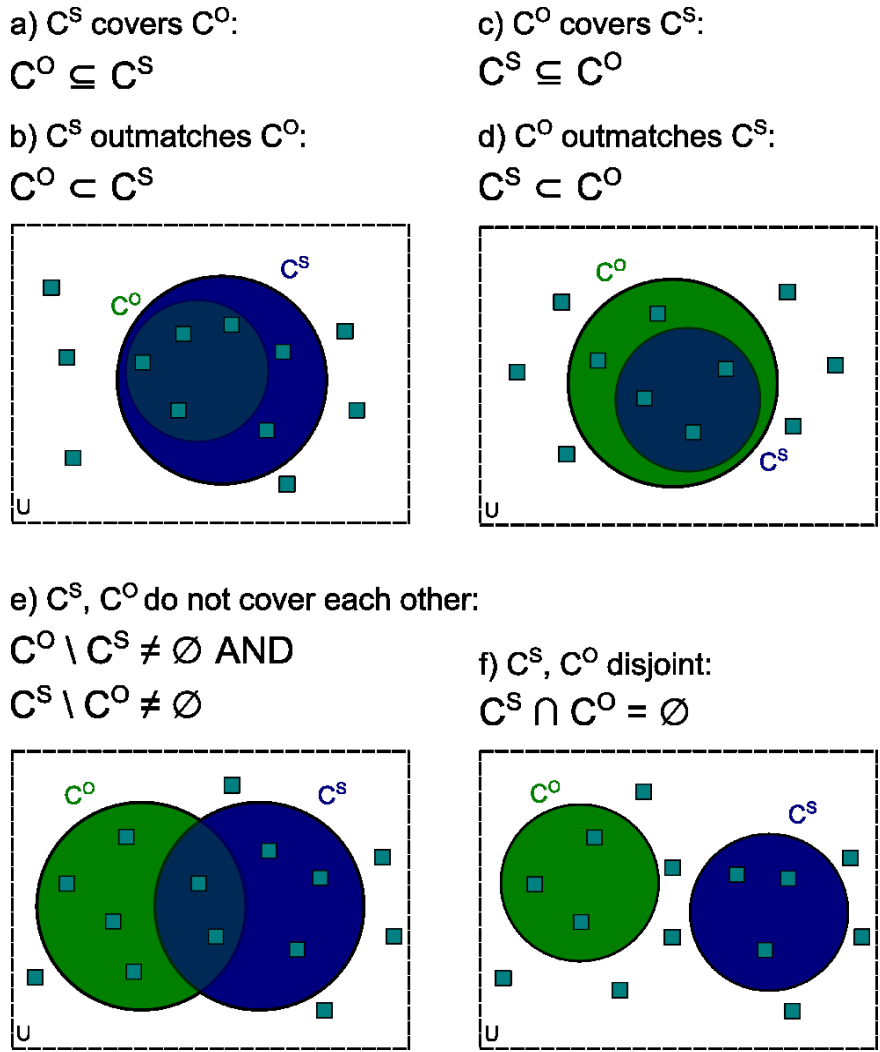

*Fig. 4: The concepts of one criterion content covering one another (a, c), one criterion content outmatching another (b, d), two criterion contents not covering one another (e), and two criterion contents disjoint (f).*

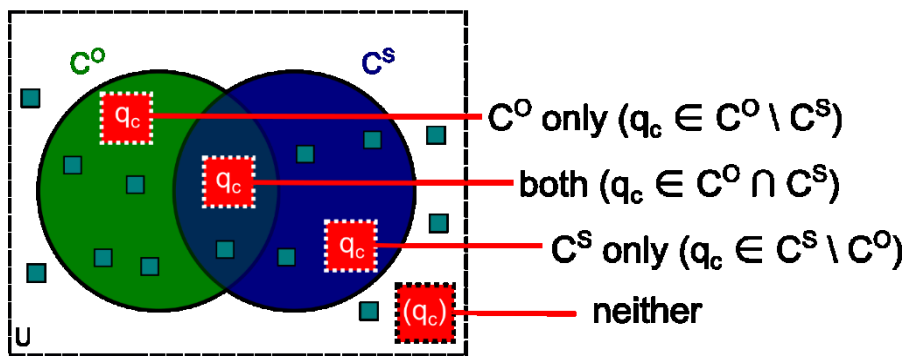

*Fig. 5: The critical cue can be part of $C^S$, $C^O$, both, or neither. It may not even exist.*

### Where is the critical cue?

It is clear that the critical cue can be contained in $C^O$ but not in $C^S$, in $C^S$ but not in $C^O$, in both, or in neither, and there is the additional possibility that it does not even exist (Fig. 5). From the representations in Figs. 4 and 5, many special cases can be constructed. For example, if $C^O$ and $C^S$ are disjoint, at most one of them can contain the critical cue, and the other one is guaranteed not to contain it. If $C^O$ contains the critical cue and $C^S$ covers $C^O$, then $C^S$ contains it too; and so on. In the following, we are looking at a number of these scenarios.

*Case A:* The critical cue is part of $C^O$ but not of $C^S$ ($q_c \in C^O \setminus C^S$). In this case, only the objective but not the subjective measure can form a dissociation with respect to the indirect task. The subjective measure either fails to ask about the critical cue, or observers execute it in a way that circumvents the use of the critical cue. For example, a subjective measure in our example experiment might ask the observer, "Please rate how clearly you perceived the prime". This instruction leaves the criterion content to the observer and is not focused specifically on the critical feature. If prime color is the critical feature, participants might perform the rating on the basis of their ability to detect the prime's presence or absence (e.g., its perceived brightness, darkness, or flicker) without looking out for its color. Later, we will see that applications of the *Perceptual Awareness Scale (PAS,* Ramsøy & Overgaard, 2004) have the problem that the unspecific questioning does not make sure that the subjective measure contains the critical cue.

If the subjective measure does not contain the critical cue, does it mean it is useless? Not at all, because it can provide valuable information beyond that needed to establish the dissociation. The paradigm case is the concomitant use of an objective prime discrimination judgment with a subjective confidence rating, as in the construction of an ROC curve.

Remember that the effectiveness of a measure jointly depends on the criterion content and on the way the cues in the criterion content are integrated. Even if $C^O$ does contain the critical cue, there is no guarantee that the observer will use it effectively or exclusively.

*Case B:* The critical cue is part of $C^S$ but not of $C^O$ ($q_c \in C^S \setminus C^O$). It is difficult to find an example where such a combination of tasks would be employed deliberately; it rather arises in cases where the objective direct measure is misspecified. One such example would be the inappropriate use of a detection task where a discrimination task would be in order. In signal detection theory, discrimination and detection can be modeled within the same decision space, but the criteria for the two tasks can be orthogonal (Macmillan, 1986): while discrimination requires a criterion that separates signal A from signal B, detection requires a criterion that separates both A and B from noise (also see Snodgrass, Bernat, & Shevrin, 2004). From this model, it would be both possible to detect a stimulus without being able to discriminate it (as in many cases of visual masking) and to discriminate it without being able to detect it (as in blindsight). Note, however, that Macmillan's (1986) model assumes that the subjective evidence for detection and discrimination is based on the same metric and can be described within a single two-dimensional space, an assumption that is called into question if both tasks are based on different criterion contents. [3]

*Case C:* The critical cue is part of both $C^O$ and $C^S$ ($q_c \in C^O \cap C^S$). This is the situation where objective and subjective direct tasks can give convergent information in the dissociation paradigm. For instance, the objective task may ask observers to discriminate the shape of the prime, and the subjective measure may ask them to rate the clarity of their shape impression. (Compare this with the foregoing example where the subjective measure was a confidence judgment.) Again, note that even if the critical cue is contained in the criterion content of a given task, there is no guarantee that an observer makes optimal use of it.

*Case D:* The critical cue is not used, $q_c \in U$, or does not exist. If the critical cue is not used in either measure, it is possible that it is principally inaccessible to the observer, or that both measures are misspecified as discussed above, or that the observer is not following instructions. A classical case where the critical cue, in all likelihood, does not exist is in research on "extrasensory perception". For instance, in an experiment where participants try to receive a telepathic image, there probably is no critical cue because there is no telepathy and thus no critical feature. The task may still be performed with some success, but only on the basis of strategic cues (e.g., educated guessing of motifs that are likely to be "transmitted").

One of the issues that can be reevaluated in light of CST is the distinction between objective and subjective thresholds. Cheesman and Merikle (1984, 1986) were the first to distinguish between objective thresholds of awareness (based on performance in objective tasks) and subjective thresholds, defined as "the prime-mask SOA at which an observer consistently claimed to detect the primes at a chance level of accuracy (p. 352)". It is generally assumed that subjective thresholds are lower than objective ones, in the sense that stronger masking is necessary to keep stimuli beneath the objective threshold (Snodgrass et al., 2004). Of course this assumption requires that the measures can be ordered in terms of their sensitivity within some kind of decision space. But from the viewpoint of CST, establishing the relative sensitivity of two measures is not trivial because it would depend jointly on the amount of overlap between the respective criterion contents, $C^O$ and $C^S$, as well as on their modes of integration. Remember that depending on how the tasks are specified, the two criterion contents may be quite dissimilar, and any attempt to order them may not be meaningful (Zehetleitner & Rausch, 2013). The only thing we can safely say at this point is that both $C^O$ and $C^S$ would need to include the critical cue, or else a dissociation with a third, indirect measure could not be established. Beyond that, however, a theory of relative sensitivity of direct measures is a task for the future.

Ironically, the successful establishment of a simple dissociation may lead to a situation where the critical cue becomes inaccessible to the observer, so that it becomes questionable whether it is still part of the criterion content of the direct task. Under visual masking, for instance, it can be very difficult or even impossible to discriminate a prime's shape or color. If that is the case in all conditions of the experiment (e.g., F. Schmidt & T. Schmidt, 2010), the observer has no choice but to resort to other cues to perform the task (e.g., auxiliary perceptual cues), working around the critical cue. It is therefore wise to mix low-visibility conditions with other conditions where the critical cue is actually helpful to ensure that the participant is still on task. Another alternative is to aim for a double dissociation where complete masking is not required (Biafora & T. Schmidt, 2020).

### Is the Perceptual Awareness Scale valid?

The *Perceptual Awareness Scale (PAS*, Ramsøy & Overgaard, 2004) is a widely used subjective measure of visual awareness and presented in its original form in Table 1. Is the scale a valid direct measure of awareness? It should be clear by now that in the dissociation paradigm, the validity of any direct measure can only be assessed in the context of the indirect task. If the indirect effect is driven by a critical feature, the direct measure must ask about that feature's representation in visual awareness, the critical cue. Otherwise, arguments for dissociation fall short because direct and indirect tasks have different objects of measurement.

*Table 1: The Perceptual Awareness Scale (Ramsøy & Overgaard, 2004). Subscript* **a** *indicates wording (highlighted by italics) that is ambiguous regarding the targeted criterion content. Subscript* **b** *indicates wording that specifically addresses decisional cues.*

| Scaling category | Category description to the observer |
|---|---|
| 1. No experience | "No impression of *the stimulus* [a]. All answers are seen as *mere guesses* [b]." |
| 2. Brief glimpse | "A feeling that *something* [a] has been shown. Not characterised by *any content* [a], and *this* [a] cannot be specified any further." |
| 3. Almost clear experience | "Ambiguous experience of *the stimulus* [a]. *Some stimulus aspects* [a] are experienced more vividly than *others* [a]. A feeling of almost being *certain about one´s answer* [b]." |
| 4. Clear experience | "Non-ambiguous experience of *the stimulus* [a]. *No doubt in one´s answer* [b]." |

Let's see if we can apply the PAS to our model experiment (Fig. 1). Do the scale categories and their respective descriptions make sense with respect to the critical features? The first thing we notice is that the rating categories are the same irrespective of the task. They do not adapt to whether we ask for the color or shape of the prime; instead, they refer rather vaguely to "the stimulus" or just "something". Moreover, the four rating categories refer to the visibility or "clarity of experience" of the prime as a whole (like in a detection task), not to the visibility of a particular critical feature. Two of the categories (2, 3) acknowledge that there might be multiple perceptual cues. Three of the categories (1, 3, 4) refer additionally to decisional cues and to the confidence in the perceptual decision (in Table 1, we marked wording that is ambiguous with respect to the criterion content or that refers to decisional cues). In sum, it seems that observers are supposed to rate two things at once: the subjective detectability (but not the discriminability) of the prime, and their own confidence in making this judgment.

Originally, Ramsøy and Overgaard (2004) used their scale in a more differentiated way. They introduced it in an experiment where the critical stimulus was one of three shapes appearing in one of three colors and at one of three locations. The scale was then applied separately to each of those stimulus features: the feature of interest was specified first and then the scale was applied specifically to it. Does this work for our model experiment? If color is the critical feature, we want to know whether observers have awareness for the distinction between red and green. If we replace "stimulus" with "color" in the scale descriptions, we again notice how fuzzy they are. "No impression of the color" and "Non-ambiguous experience of the color" are still reasonably clear. But what about "A feeling that something colored has been shown. Not characterized by any content, and this cannot be specified any further"? What would the "content" of the colored thing mean here -- the specific hues of red or green? And what is the ominous "this" that cannot be specified any further -- the "content" or the "characterization"? In fact, the two middle categories are formulated in a way that it is difficult to reconcile them with feature discrimination; they are clearly designed with detection in mind. Because the dissociation paradigm almost invariably employs an indirect task based on discrimination and not detection, and because those two types of task may be based on orthogonal decision criteria (Macmillan, 1986), the PAS is generally not a suitable choice.

Let's switch perspective and ask what a dissociation experiment would look like for which the PAS would be a good choice. Because the direct measure focuses on detectability (plus confidence), the indirect task would have to be a detection task as well. The indirect effect would thus depend on the presence or absence of the critical feature, not its identity. The labeling of one rating category as "brief glimpse" further excludes tasks where the critical stimulus is presented for prolonged times, as in binocular rivalry, continuous flash suppression, or some inattention paradigms. It confines the PAS to experiments with briefly flashed stimuli, like masked priming or the attentional blink paradigm. [4]

**How well do you know your direct measure? A checklist.**

From the foregoing, it should be obvious that the choice and construction of a suitable direct measure requires a lot of consideration, both on the theoretical and on the practical side. Table 2 provides a checklist for properties of direct measures that integrates many of the issues discussed in this paper, plus some practical issues that frequently arise in the measuring and testing process.

**Claims of measurement properties and what they require**

The literature is full of claims concerning the measurement properties of various direct measures as measures of visual awareness, for instance their validity, exhaustiveness, or exclusiveness. If we translate the expression "Measure $M$ is a valid measure of visual awareness" into "Measure $M$'s criterion content $C^M$ contains the critical cue", we can specify the assumptions the respective claim has to meet (Table 3).

*Table 2: A checklist for properties of direct measures.*

| |
| --- |
| What is the critical feature that drives the indirect effect? |
| Do you want to establish a dissociation between the direct and an indirect measure? If so, do you require the direct measure to be exhaustive for conscious information (as for simple dissociations) or merely monotonic (as for double dissociations)? |
| Is your direct measure objective, subjective, or hybrid? |
| Is there a psychophysical model underlying your measure (e.g., a psychometric, signal-detection, or threshold model)? Do you have an idea about the decision space involved, the nature of the decision axis, the placement of criteria, etc.? Can sensitivity be separated from decision bias? What scale level do you assume (nominal, ordinal, interval, ratio scale)? |
| Does the direct measure explicitly ask for the critical cue? If not, are you confident that the critical cue is contained in the criterion content? Which other cues may be theoretically relevant and should be captured by the measure? |
| Does your measure avoid other forms of D-I mismatch with the indirect measure (discrepancies in stimuli, responses, S-R mapping)? |

| What cues do you expect to be in the criterion content besides the critical cue? Could an observer use the measure in a different way than intended (i.e., based on undesirable cues)? |
| :--- |
| Is your measure reliable within single observers? How precise is the measurement in terms of standard errors in each participant and condition? (As a rule of thumb, try 100 repetitions per observer and condition.) |
| Is your measure consistent across observers, or are there qualitative differences from person to person? Is it safe to average your measure across observers, or does each observer have to be considered separately? How do you plan to analyze your data in those circumstances? |
| Is your direct task too difficult? Does it discourage observers from trying to evaluate a stimulus they are unable to access? Do you need to train your observers to perform the task in a specific way? |
| Do you avoid sampling artifacts such as selective analysis of participants, isolation of zero visibility ratings as "unconscious", etc.? |
| If you apply several direct measures, how are they related? Are they supposed to converge (with overlapping criterion contents) or to give independent information (with disjoint criterion contents)? |
| Are there any dissociations among the direct measures? Are there double dissociations? |
| Do you use several measures concurrently on the same trial? If so, how large is the working memory load of the multitask, how strong is the reliance on memory representations, how large are possible interference effects between subtasks? Do the awareness measures interfere with the indirect effect (e.g., by prolonging response times or changing the structure of a priming effect)? |
| Do you use several measures in different blocks or sessions? If so, how does the order of the tasks affect the measurement? Are you taking care to test your hypothesis conservatively? |

*Table 3. Frequent claims about properties a direct measure M, the assumptions about the criterion content $C^M$ implied by those claims, and possible counterarguments against those claims.*

| Claim: | Assumptions about criterion content implied by the claim: | Counterargument: |
| :--- | :--- | :--- |
| $M$ is a valid measure of awareness. | Weak assumption:<br>$q_c \in C^M$ | Difficult to counter even for obviously misspecified tasks. |
| $M$ is an exclusive measure of awareness. | Strong assumption:<br>$C^M = \{q_c\}$ | Show that $M$ is sensitive to parameters other than the critical stimulus. |
| Only class $S$ of subjective tasks can measure awareness. | Very strong assumption:<br>For all measures $N \notin S$, $q_c \notin C^N$ | Construct objective analogs to the subjective tasks (often possible). |
| Only class $O$ of objective tasks can measure awareness. | Very strong assumption:<br>For all measures $N \notin O$, $q_c \notin C^N$ | Construct subjective analogs to the objective tasks (usually possible). |
| Only $M$ can measure awareness. | Prohibitive assumption:<br>For all measures $N \neq M$, $q_c \notin C^N$ | Show that measures other than $M$ can ask for the critical feature. |
| $M$ is an exhaustively valid measure of awareness. | Strong assumption:<br>$C^M$ includes all theoretically relevant cues | Show that $M$ fails to respond to some theoretically relevant cue that another measure can respond to. |
| $M$ is an exhaustively reliable measure of awareness. | Prohibitive assumption:<br>$M$ is exhaustively valid and an exhaustive integrator of all theoretically relevant cues | Show that $M$ has reliability < 1 or an appreciable standard error. |

The claim that a direct measure does at least have some validity is difficult to dismiss. Given our definition of validity, the claim only requires that $q_c$ be an element of the criterion content of the task. Even if a task is grossly misspecified, it is possible that participants spontaneously use the critical cue anyway. The PAS scale, for instance, was designed primarily with quickly presented stimuli in mind. If it is instead applied to temporally extended stimuli, like in rivalry paradigms, observers that have awareness of the critical feature may spontaneously rate their experience on their own internal four-point scale and map it to the PAS categories, even if the wording does not fit (as anticipated in Kahneman, 1968).

The claim that a direct measure is an exclusive measure of awareness is only true if the criterion content consists solely of the critical cue and no other sources of information are used (not even those that are correlated with $q_c$). This is a strong claim that can be countered empirically by showing that task performance is influenced by factors other than the critical stimulus feature, e.g., by inducing different response strategies to show that strategic cues are being used on top of the critical cue.

The claim that only one specific task can measure awareness is even stronger because it is only true if every other measure's criterion content is devoid of the critical cue. This is an implausible assumption because other direct measures will be correlated with the task at hand under parametric variations of the critical feature (e.g., increasing color contrast would not only increase color discrimination performance but also confidence ratings or clarity ratings). And of course, there are usually alternative measures that also address the critical cue directly.

The claim that only one class of measures (e.g., only subjective ones or only objective ones) can measure awareness is frequently encountered in the literature, often with the pretension that one or the other class be a "gold standard" in measuring awareness. But such a claim is only true if for all measures not contained in that class, the criterion content is devoid of the critical cue. This is implausible for two reasons. First, there are many situations where objective and subjective measure are highly correlated (e.g., Peremen & Lamy, 2014), and it is difficult to dismiss the possibility that they both use the critical cue. Second, it is often possible to find pairs of objective and subjective tasks that directly ask for the critical cue or feature. For instance, the objective task "Determine whether the prime was a square or a diamond" can easily be translated into a subjective task, "Could you see whether the prime was a square or a diamond? Answer yes or no." If objective and subjective tasks both explicitly ask about the critical cue, there is a strong possibility that it is used in both tasks.

Sometimes a measure is proposed to be exhaustive. We saw above that exhaustiveness has two aspects: one concerns the reliability of the measurement (whether the direct measure is an exhaustive integrator, i.e., a strictly monotonic function of the cues in its criterion content; T. Schmidt & Vorberg, 2006) and the other one concerns its validity (whether the criterion content comprises all theoretically relevant cues). Exhaustive reliability is usually out of the question because any empirical psychophysical measure will have a reliability clearly < 1 as well as an appreciable standard error, and for these reasons alone must be expected to violate strict monotonicity. Exhaustive validity, on the other hand, requires the slightly less extreme assumption that the criterion content contains all relevant cues and that none of them remains unused. A claim that a measure is exhaustively valid can be countered by demonstrating that there is some theoretically relevant cue that the measure does not respond to, even though a rival measure could. Even though this result could also occur when the measure in question does use all relevant cues and is just not optimally integrated, it calls the claim into doubt.

As an example, in Koster et al.'s (2020) study the data indicates that the objective measure (prime discrimination) is not exhaustively valid. If it were, its criterion content would include the perceived rotation between prime and mask because this rotation predicts the congruency of prime and mask: for instance, if you see a square target preceded by a rotating motion, you can infer that the prime has probably been a diamond. Objective discrimination performance should then increase, not decrease, with SOA, because perceived rotation increases as well. This implies that $C^O$ fails to cover the $C^S$ of the rotation measure: There is at least one cue in that subjective measure's criterion content that is not utilized in the objective measure.

**Open-feature indirect tasks and invalid indirect measures**

There are indirect tasks that do not generate a well-defined critical feature in the first place. The most important examples include indirect effects that do not depend on the identity of a prime, but on its presence or absence (implicit detection tasks). Because detection can occur on the basis of *any* stimulus feature (it is based on the disjunction of all features), it is not clear from the task whether any one of them is critical, or which one is (see Wilken & Ma, 2004, for models of change detection in a disjunction of features). Other indirect effects may be driven by same-different distinctions or oddity detection (e.g., mismatch negativity, oddball tasks). For example, van Opstal, Gevers, Osman, and Verguts (2010) show that when observers make same-different judgments on a pair of target stimuli, they are primed by same-different relations in a masked pair of primes, even though the primes and targets come from separate stimulus domains. In such a task, it is difficult to pinpoint the critical feature.

*Open-feature tasks* are tasks that deliberately leave the choice of criterion content to the observer. This can be a great advantage: for instance, animals or small children can indicate whether two stimuli are the same or different even though they are unable to verbalize the difference (for instance, they may look preferentially at a new or mismatching stimulus and thus indicate that they have processed the difference). But for the dissociation paradigm, open-feature tasks provide a great challenge. Instead of fulfilling their "anchoring" function of providing a single critical feature, they are based on a set of possible features that all might drive the indirect effect. In consequence, these indirect tasks are based on some criterion content of their own, and the set of cues in that criterion content is usually not precisely known. If that is the case, it is difficult to find direct tasks that can provide a valid comparison, and any apparent "dissociation" is easily one between apples and oranges. Open features can greatly complicate the formal analysis of the dissociation paradigm: They essentially turn Fig. 3 into a display of three overlapping sets and increase the number of special cases to be considered.

Such a mismatch, of course, can occur with more defined indirect measures as well. Consider the problem of demonstrating affective priming by schematic face stimuli (smiley and frowny faces; e.g., Fenske & Eastwood, 2003; but compare F. Schmidt & T. Schmidt, 2013). In that field, authors usually assume that the priming effect is based on an affective response to the prime or at least on its semantic processing, but Horstmann, Borgstedt, and Heumann (2006) argue that it is driven primarily by low-level visual features in the stimuli. In such a situation, a direct task asking for affective evaluation of the prime

(e.g., rating its friendliness on a scale from -3 to 3) would be a mismatch to the indirect task. A direct task directly asking for the presence of low-level visual features, on the other hand, might be a better match. Unfortunately, it would no longer be addressing the original research idea because the indirect task is invalid to begin with (confounded by low-level features).

## Double dissociations among direct measures: moving beyond the classical dissociation paradigm

The dissociation paradigm in its classical form is based on the comparison of one direct and one indirect measure. If those two measures form a double dissociation (one increasing under experimental manipulation, the other decreasing), then we can dismiss the possibility that both measures are monotonic functions of the same single source of information (T. Schmidt & Vorberg, 2006; Biafora & T. Schmidt, 2020). In particular, they cannot both be based on a single source of *conscious* information, so that there must be a second information source dissociable from it. Of course, the same logic applies when we compare several direct measures. Many measures have been proposed that are all supposed to measure awareness of the prime. But if there are double dissociations among those measures, it follows that they cannot all measure the same unitary source of information.

In the following, we explore the consequences of double dissociations among a set of direct measures. We assume that all cues and measures are scaled with the same polarity, such that larger values indicate greater evidence for the feature in question. Following T. Schmidt and Vorberg (2006), we define dissociations by comparing measures under pairs of experimental conditions. To simplify matters, we use the symbols, <<, >>, and == to indicate that measures obtained under two experimental conditions are unequivocally different or similar, for instance because they passed a statistical or numerical criterion (leaving aside the statistical issues).

*Definitions (iv).* Let $A_i$ and $B_i$ denote two measures $A$, $B$ with criterion contents $C^A$, $C^B$, observed under two experimental conditions $i$, $i \in \{1, 2\}$. Assume that $A$ and $B$ are scaled with the same polarity. Then $A$ and $B$ form a *simple dissociation* if $A_1 << A_2$ and $B_1 == B_2$ (or vice versa), a *double* *dissociation* if $A_1 \gg A_2$ and $B_1 \ll B_2$ (or vice versa), and an *association* if either $A_1 \ll A_2$ and $B_1 \ll B_2$ or $A_1 \gg A_2$ and $B_1 \gg B_2$.

We next prove that two double-dissociated measures cannot measure the same unitary content. The proof closely follows the one in T. Schmidt and Vorberg (2006).

*Proposition 1 ("no single source for double dissociations"):* Assume two measures $A$, $B$ with criterion contents $C^A$, $C^B$, that are both monotonic integrators, are scaled with the same polarity, and are observed under experimental conditions $i$, $i \in \{1, 2\}$. Then a double dissociation between $A$ and $B$ rules out that both criterion contents consist of the same single cue, $q$.

*Proof:* Suppose that $A_1 \ll A_2$ while $B_1 \gg B_2$ (the proof for the reverse case is analogous). We show that the postulate $C^A = C^B = \{q\}$ leads to a contradiction. By this postulate and the assumption of monotonic integration, both $A$ and $B$ are monotonic functions of $q$ only. The observation that $A_1(q) \ll A_2(q)$ thus implies that $q$'s value has increased from condition 1 to condition 2. At the same time, the observation that $B_1(q) \gg B_2(q)$ implies that $q$'s value has decreased in value from condition 1 to condition 2, which completes the contradiction.          □

Consider again the eight measures included in Koster et al.'s (2020) study. We start by theorizing that they are all measures of the same unitary perceptual content, "awareness of the prime". In other words, we postulate that for each measure $M_i$, the criterion content $C_i^{M_i} = \{q_a\}$, where $q_a$ is awareness of the prime. This postulate runs into trouble because there are double dissociations between some of the measures, which implies that they cannot all measure the same thing. This has far-reaching consequences for other potential measures, even those not included in Koster et al.'s set. As soon as two measures form a double dissociation, dissociative relations can spread across the entire network of potential measures. To see this, consider a direct measure $M_{inc}$ that clearly increases under experimental manipulations, and another direct measure $M_{dec}$ that clearly decreases. The two form a double dissociation with respect to awareness of the prime, which we write as $DD(M_{inc}, M_{dec})$. But any other measure $M_i$ that clearly increases or decreases under the manipulation will be double-dissociated with either $M_{inc}$ or $M_{dec}$: either $DD(M_i, M_{inc})$ or $DD(M_i, M_{dec})$. It becomes clear that double dissociations are *contagious*: as soon as there is even one in a set of possible measures of the same perceptual content, there will likely be others.

**Consequences for theory-building: "Explaining the gradient".**

We are now ready to leave the narrow confines of the dissociation paradigm by giving up the distinction between direct and indirect measures. The privileged roles of the indirect task, the critical feature, and the critical cue all fall away. What remains is a large set of possible measures with different criterion contents, different modes of integration, different measurement properties, and different measurement objectives. Some may be direct, others indirect; some objective, others subjective (Fig. 6). Whenever any two of them become double-dissociated, their criterion contents may overlap but cannot be constricted to a single perceptual content.

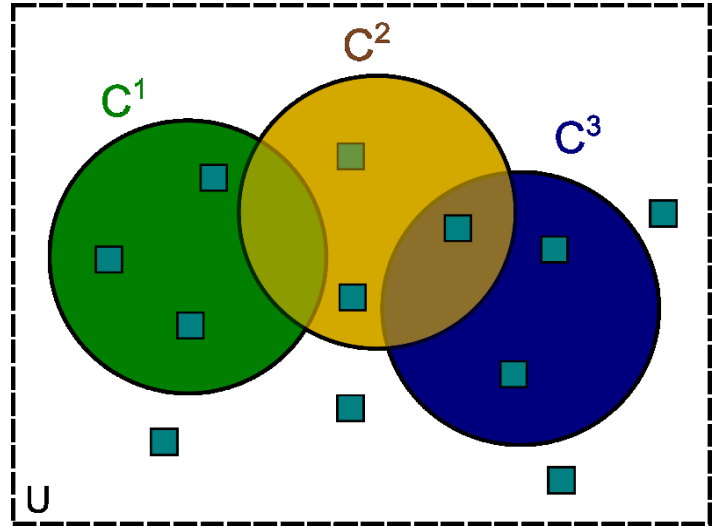

*Fig. 6. When the distinction between direct and indirect measures is given up, what remains is a set of measures whose criterion contents may overlap in various ways. Double dissociations between any two of them imply that their criterion contents cannot be restricted to the same single cue.*

The empirical observation of double dissociations among direct measures (Koster et al., 2020) has immediate consequences for any theory of consciousness. In particular, any theory that explains "consciousness" by a single monotonic process or mechanism is challenged by double dissociations between direct measures. The proof is practically identical to the previous one:

*Proposition 2 ("no simple theory for double dissociations")*: Assume two direct measures $C_i$, $D_i$ that are scaled with the same polarity and are observed under experimental conditions $i$, $i \in \{1, 2\}$. Assume a theory $T$ that explains variations in $C$ and $D$ as a monotonic function of a single process $p$, $C = f(p)$ and $D = g(p)$, such that $f(p') \leq f(p)$ and $g(p') \leq g(p)$ for any $p' \leq p$. Then $T$ is falsified by a double dissociation between $C$ and $D$.

*Proof (Schmidt & Biafora, 2022):* Suppose that $C_1 << C_2$ while $D_1 >> D_2$ (the proof for the reverse case is analogous). We show that the postulate $C = f(p)$ and $D = g(p)$ leads to a contradiction. By assumption of monotonicity of $f$ and $g$, both $C$ and $D$ are monotonic functions of $p$ only. The observation that $C_1(p) << C_2(p)$ thus implies that $p$'s value has increased from condition 1 to condition 2. At the same time, the observation that $D_1(p) >> D_2(p)$ implies that $p$'s value has decreased from condition 1 to condition 2, which completes the contradiction. Therefore $C$ and $D$ cannot both be monotonic functions of $p$, and theory T is falsified. □

If several direct measures are available simultaneously, we call this set a *gradient* (Schmidt & Biafora, 2022). Here, we are especially interested in gradients that contain at least one double dissociation between direct measures, such that one of them increases across stimulus conditions while another of the same polarity decreases. Proposition 2 immediately generalizes to gradients of multiple direct measures by requiring any theory of visual awareness to explain the gradient of awareness measures. If a gradient contains at least one double dissociation, any theory explaining "consciousness" by means of a monotonic function of a single process is falsified:

*Propositon 3 ("explaining the gradient")*: Suppose a set of direct measures $D_j$, $j = \{1, 2,…\}$, that are scaled with the same polarity and where at least two of the measures form a double dissociation. Let $T$ be a theory that explains variations in $D_j$ as a monotonic function $f_j$ of a single process $p$, $D_j = f_j(p)$, such that $f_j(p') \leq f_j(p)$ for all $p' \leq p$. Then $T$ is falsified.

*Proof:* Because the set of direct measures contains at least one double dissociation by definition, Proposition 2 applies. □

**General discussion**

Cue Set Theory is a theory of measurement. Just as an empirical theory can be judged by the data it can explain and predict, a theory of measurement can be evaluated by its ability to clarify concepts, clear up misunderstandings, explain and predict methodological difficulties, and ultimately improve measurement tools. We believe that CST contributes to the clarification of concepts by elucidating the idea of criterion content, its variation across awareness measures, and the crucial role of the critical cue as a pivot between direct and indirect measures. CST also clarifies the concept of exhaustiveness as having a validity as well as a reliability aspect – a distinction that was not clear to us before we attempted to formalize our theory. We further hope that CST will help clear up some fundamental misunderstandings, e.g., the one that in order to measure subjective awareness, the direct measures must likewise be subjective. Indeed, the sometimes fierce battles between proponents of subjective and objective measures should largely be settled by the recognition that such measures may have overlapping but nonidentical criterion contents, that either may give invaluable information the other one could not provide, and that both types of measures can comfortably be united, e.g., in a simple bipolar rating scale. CST therefore has the potential not only of improving the quality of awareness measures, but also to specify the scopes and limitations of each such measure (e.g., in our critique of the Perceptual Awareness Scale). Finally, even a theory of measurement may have some capacity for empirical predictions. Specifically, we expect that double dissociations between measures of awareness will be abundant, will continue to be discovered, and will require more specific and more refined theoretical explanations. In the course of that, we expect that research into cognition without awareness will increasingly be viewed as a study of task dissociations. Nevertheless, the dissociation paradigm will remain instrumental for providing the database for such a research project if applied in a straightforward, principled way.

CST expands Kahneman's (1968) notion of criterion content to sets of cues. Importantly, these cues are not necessarily perceptual – they may include feedback from the motor system, feedback from the decision-making process, and strategic knowledge of the task. In this regard, CST differs from other multifeatural conceptions of awareness that only consider perceptual information at different levels of an assumed processing hierarchy (Fazekas & Overgaard, 2018; Kouider, de Gardelle, Sackur, & Dupoux, 2010). The most important property of the dissociation paradigm according to CST is that the indirect task implies a critical feature that anchors the critical cue -- the critical aspect of perceptual experience that the direct measures are supposed to capture. Direct measures whose criterion contents do not contain this critical cue are often valuable and informative in their own right but provide no foundation for establishing a dissociation with the indirect measure. They essentially try to compare their own apples with the indirect measure's oranges (Erdelyi, 1986). Likewise, unspecific measures that do not focus explicitly on the critical cue, like the Perceptual Awareness Scale, do not provide a solid basis to argue for dissociation from the indirect measure. Fortunately, it is often straightforward to identify the critical feature and to construct direct measures targeting it, both objective and subjective ones.

CST explicitly acknowledges that criterion content may vary between observers – perhaps because of idiosyncratic differences in their perceptual systems (e.g., in the time course of visual masking; Albrecht & Mattler, 2016), because of different strategies in forming the criterion content, but also because of different ways of integrating the available cues (Bernstein, Fisicaro, & Fox, 1976; Jannati & DiLollo, 2012; Ventura, 1980). Of course, such idiosyncrasy complicates measurement as well as the interpretation of measures. One course of action is to use direct measures that explicitly ask for a particular content, and to train observers to report only on that content. In our opinion, it is furthermore essential to set up experiments in such a way that individual data patterns can be evaluated reliably. This is why we prefer a small number of trained observers performing many trials (generally, several sessions) to a large group of observers performing only few trials. We are therefore following the psychophysical measurement standard now discussed under the label "small-N design" (Smith & Little, 2018; see Arend & Schäfer, 2019, and Baker et al., 2021, for demonstrations and easy calculations of adequate statistical power in such designs). Note that it is never advisable to average across observers with qualitatively different data patterns. [5]

CST also provides a new justification for employing different direct measures in the same paradigm. Once there is at least one direct measure that is reasonably valid in capturing the critical cue, other measures can focus on different facets of measurement. Because these additional measures do not need to utilize the critical cue, their criterion contents are free to give separate information about the observer's performance or experience. And indeed, if no measure at all is singled out to anchor the critical feature, all that remains are multiple facets of measurement that can be compared for their properties. For example, Zehetleitner and Rausch (2013) show that ratings of decision accuracy can outperform stimulus ratings. Koster et al. (2020) show that performance in the objective discrimination task can be low even in the presence of rich subjective perception of other aspects of the critical stimulus, and that those subjective cues can be dissociated among each other. Similarly, Maniscalco, Peters, and Lau (2016) demonstrate that double dissociations can occur between two direct measures (an objective measure of performance, $d'$, and a subjective measure of confidence, meta-$d'$) and how such a dissociation is predicted by signal detection theory.

Giving up the distinction between direct and indirect measures takes care of a fundamental puzzle in the history of consciousness research: What is used as a direct measure of awareness in one study may be used as an indirect measure of unconscious processing in another (Timmermans & Cleeremans, 2015). For instance, Peirce and Jastrow (1885) argued that their participants could successfully discriminate between two objects even though they indicated that they had no confidence in their decisions. Similarly, Sidis (1898) tried to create stimulus conditions such that participants reported being unable to detect the stimuli and yet showed some ability to discriminate between them, and concluded that discrimination was based on unconscious perception (a recent paper from Stein and Peelen, 2021, uses the same argument). From the point of view of the later measurement tradition in unconscious perception, those authors used measures that were all indicators of visual awareness and demonstrated dissociations between them. From the perspective of CST, however, there is no contradiction because different direct measures are assumed to be based on different criterion contents, which allows for dissociations between different facets of awareness.

From such considerations, Koster et al. (2020) draw the following conclusions:

"[…] subjective experience has to be conceived as a multidimensional pattern of experiences. It is important to note that this finding casts doubt on all attempts to measure visual awareness in a single univariate measure because some other aspects of visual experience might always vary in opposite ways across a given parameter such as SOA. In consequence, the idea of an exhaustive measure or a gold standard for measuring consciousness appears simplistic." (p. 20)

The philosopher Elizabeth Irvine (2017) comes to a similar conclusion. In a paper entitled "Explaining What?", she distinguishes between the concepts of "Konsciousness" (with a capital "K") and "schmonciousness". Believers in Konsciousness have a hard, monolithic concept of what they want to explain, "a single, coherent and unitary explanatory target" that may find explanation in a single sweeping theory. Believers in "schmonciousness", on the other hand, have a much more modest concept: they assume that the term "consciousness" is still volatile and maybe even prescientific, and that it may disintegrate into the study of many more specific aspects:

"Rather than keep trying (and failing) to identify which state or process consciousness *really* is, the idea is to accept the fragmentation […]. […T]rying to explain consciousness with a single materialist blow is just as confused as trying to explain intelligence […], health or happiness by pointing to a single mechanism, gene, or causal factor." (p. 9)

CST leads to a similar conclusion. Our Propositions 1-3 state that theories explaining the entirety of consciousness out of a single monotonic process are falsified by double dissociations among measures of awareness. This leads to a simple demand that can be placed on any theory aiming to explain consciousness. We call this demand "Explaining the gradient" (T. Schmidt & Biafora, 2022). We define a *gradient* as a set of measures responding to specific changes in experimental conditions; e.g., Figure 2 shows a gradient that consists of a small set of direct measures in response to a variation in prime-target SOA. When experimental conditions are varied, the gradient may respond in complex ways: some measures may increase with parameter changes, others decrease; some may be u-shaped, others invariant; some may respond to some experimental variations but not others. Convincing theories of visual awareness should aim at explaining such gradients, at least for some direct measures and some experimental conditions at a time (see Doerig, Schurger, & Herzog, 2020, for further criteria that could be applied to such theories).

Explaining the gradient requires a theory that is sufficiently specific about the facets of awareness involved (Klein & Hohwy, 2015). If visual awareness consists in a multi-dimensional pattern of dissociable cues, each of those cues requires sophisticated measurement, explanation, and theorizing. A theory trying to explain the simultaneous experience of color and motion in a masked stimulus must therefore involve a theory of color and motion before it can begin to explain why these impressions are conscious. Such a theory of consciousness is not in sight. The most prominent current theories attempt to explain "consciousness" out of a single process: For instance, Global Workspace Theory postulates a widespread "ignition" in neural activity (Baars, 1993, 2013, Dehaene & Naccache, 2001), and Integrated Information Theory postulates that consciousness is a consequence of the amount of "integrated information", $\Phi$ (Tononi & Edelman, 1998, Tononi, 2004, but see Oizumi, Albantakis, & Tononi, 2014, for a formulation of the theory that seems to allow for multiple $\Phi_i$). These theories have in common that their explanatory process is strictly unidimensional. Because they are trying to specify the neural correlate of a unitary process of consciousness, they are not able to explain why one facet of visual awareness increases while another one decreases: they fail to "explain the gradient" of the experiment. Ultimately it is the dynamics of those facets that need to be explained by any fully-developed theory of visual awareness. We hope that Cue Set Theory will help transform the field of consciousness research to a detailed, sophisticated study of task dissociations among direct and indirect measures, not merely by "accepting the fragmentation" (Irvine, 2017), but by appreciating the fascinating kaleidoscopic nature of conscious and unconscious vision.

### Footnotes

[1] In this paper, we avoid the term "subliminal perception" because the concept of a "limen" or "threshold" is meaningful only in the context of a concrete psychophysical threshold model (for introductions to psychophysical models, see Gescheider, 1997, and Macmillan & Creelman, 2005). The purely metaphorical use of the term (e.g., calling a stimulus "subliminal", "passing the threshold to consciousness", etc.) continues to be a major source of confusion because it intuitively suggests a single-high-threshold model that has largely been discredited by empirical data (Wixted, 2020).

[2] The critical feature is not always defined by a single physical property. For instance, when the task is to distinguish male from female face targets, the critical feature is indeed the sex of the photographed person, even though the classification probably involves multiple facial features. The example becomes even clearer when the task is to distinguish male from female *names*, where the feature is the semantic class membership of the name. Ultimately, it will be the experiment's programming code that defines the critical feature.

[3] Just as the criteria for discrimination and detection are different, so are the criteria for yes-no discrimination, two-alternative forced choice, same-different tasks, ABX tasks, matching-to-sample, and many other variants (Macmillan & Creelman, 2005). They are suitable indirect tasks for the dissociation paradigm only if they have a clearly defined critical feature.

[4] Sandberg and Overgaard (2015) state that the specific wordings of the rating categories can be adjusted rather freely and still yield a PAS. In their view, this even holds for changing the number of categories in the scale as long as they are in a 1:1 correspondence with the observers' subjective states. From a psychometric point of view, we find this view problematic, as the validity and reliability of any scale will respond strongly to such momentous changes. A widespread current practice seems to be that researchers start from the original PAS categories to construct their own custom-made scales, to which they still refer to as "PAS" even when there are strong modifications.

[5] Other harmful practices should be avoided, too (F. Schmidt, Haberkamp, & T. Schmidt, 2011). One is to use weak statistical tests of direct measures (e.g., a between-subjects *t*-test of $d'$ against zero where sensitivity is averaged across a small number of participants) because they are extremely lenient towards accepting the null hypothesis of processing without awareness. A powerful alternative is to calculate a $\chi^2$ test of hits, misses, false alarms and correct rejections for each of the $N$ observers, and then to cumulate the $\chi^2$ values to test them against a $\chi^2$ distribution with $N$ degrees of freedom (Vorberg et al., 2003). This test is powerful and strict because it employs a cumulation of $N$ within-subject tests. – Another poor practice is to use only a fragment of the experimental time on the direct task – if anything, the direct test usually requires at least as many trials as the indirect one, and often more. A highly problematic but still popular practice is to sort trials *post hoc* on the basis of concurrent visibility ratings in order to restrict testing to trials in the zero-visibility category (as proposed by Bachmann & Francis, 2014, and van den Bussche et al., 2013, among many others), or to even exclude observers or trials for whom direct performance exceeded some criterion. Such procedures capitalize on chance factors by distorting the samples, by artifactually creating samples of participants with zero performance in the direct task, and by creating spurious "invisible" trials even if masking is ineffective (T. Schmidt, 2015). They invariably lead to regression to the mean, thus overestimating or even fabricating the dissociation (Shanks, 2017).

### Preprint note

This manuscript is currently under review and is still subject to change.

### Open practices statement

Data, materials and analyses for our datasets are available from the authors upon request.

### Author note

This manuscript is a preprint that is expected to change in the process of publication. Correspondence may be sent to T.S. (thomas.schmidt@sowi.uni-kl.de) or M.B. (melanie.biafora@sowi.uni-kl.de). We thank Sven Panis and Max Wolkersdorfer for helpful comments on previous versions, and Anabel Weiß and Laura Wetz for checking the reference list. Numerous reviewers have contributed to this work, and we thank them all. We have taken care to include as many of their suggestions as we could to improve the manuscript. Special thanks go to Wolfgang Lenski for improving our manuscript from a perspective of logic and epistemology.